\documentclass[final,5p,times,twocolumn,preprint,sort&compress]{elsarticle}
\usepackage{bm} 
\usepackage{graphicx}
\usepackage{epstopdf}
\usepackage{hyperref}
\usepackage{amsmath,amssymb,amsfonts,latexsym}
\newcommand{\lambdaDp}{{\lambda'}_{\!\! D}}
\newcommand{\lambdanp}{{\lambda'}_{\!\! n}}

\allowdisplaybreaks
\begin{document}
\begin{frontmatter}
\title{Neutron spin structure from polarized deuteron DIS with proton tagging} 
\tnotetext[t1]{JLAB Report JLAB-THY-19-2967}
\author[c]{W.~Cosyn}
\address[c]{Department of Physics and Astronomy, Ghent University, Proeftuinstraat 86, B9000 Ghent, Belgium}
\ead{wim.cosyn@ugent.be}
\author[w]{C.~Weiss}
\address[w]{Theory Center, Jefferson Lab, Newport News, VA 23606, USA}
\ead{weiss@jlab.org}
\begin{abstract}
Polarized electron-deuteron deep-inelastic scattering (DIS) with detection of the spectator proton
(``tagged DIS'') enables measurements of neutron spin structure with maximal control of nuclear effects. 
We calculate the longitudinal spin asymmetries in polarized tagged DIS using methods of light-front
nuclear structure and study their dependence on the measured proton momentum. Asymmetries can be formed with 
all three deuteron spin states ($\pm 1, 0$) or the two maximum-spin states only ($\pm 1$, involving 
tensor polarization). The proton momentum dependence can be used to select pure S-wave configurations 
in the deuteron and eliminate D-wave depolarization (transverse momenta $p_{pT} \lesssim$ 100 MeV).
Free neutron spin structure can be extracted model-independently through pole extrapolation
of the tagged asymmetries. Such measurements could be performed at a future electron-ion collider (EIC) 
with polarized deuteron beams and forward proton detectors.
\end{abstract}
\end{frontmatter}
\section{Introduction}
Nucleon spin structure studies require measurements of polarized deep-inelastic lepton scattering (DIS)
on both the proton and the neutron; see
Refs.~\cite{Anselmino:1994gn,Kuhn:2008sy,Aidala:2012mv} for a review.
Proton and neutron data together are needed to determine the flavor 
composition of the quark helicity distributions, to separate singlet and nonsinglet structures in the
analysis of scale dependence (QCD evolution, higher-twist effects), and to evaluate the Bjorken sum rule. 
The neutron spin structure functions are extracted from DIS measurements on polarized light nuclei 
(deuteron $^2$H $\equiv D$, $^3$He). The procedure must account for nuclear effects such as 
nucleon spin depolarization, motion of the bound nucleons, non-nucleonic degrees of freedom
(e.g.\ $\Delta$ isobars in $^3$He), and nuclear shadowing and 
antishadowing \cite{Frankfurt:1988nt,CiofidegliAtti:1993zs,Melnitchouk:1994tx,Kulagin:1994cj,Piller:1995mf,%
Frankfurt:1996nf,Bissey:2001cw,Ethier:2013hna}. 
The theoretical treatment of these effects is complicated by the fact that 
they depend strongly on the nuclear configurations present during the high-energy process. 
In inclusive DIS measurements one accounts for the effects by modeling them in all possible 
configurations and summing over them, which results in a 
significant theoretical uncertainty. In view of the experimental precision achievable with a future 
electron-ion collider (EIC) \cite{Boer:2011fh,Accardi:2012qut}, it is necessary to consider new types 
of measurements that reduce the theoretical uncertainty in neutron spin structure extraction.

DIS on the deuteron with detection of a proton in the nuclear fragmentation region (``tagged DIS''),
$e + D \rightarrow  e' + X + p$, represents a unique method for neutron structure measurements.
The deuteron wave function in nucleonic degrees of freedom ($pn$) is simple and known well up to
momenta $\sim$ 300 MeV; $\Delta$ isobars are suppressed in the isospin-0 system \cite{Frankfurt:1981mk}.
The detection of the proton identifies events with active neutron and eliminates dilution.
The measurement of the proton momentum fixes the nuclear configuration during the high-energy process 
and enables a differential treatment of nuclear effects. Extrapolation of the measured proton momentum 
dependence to the on-shell point eliminates nuclear initial-state modifications and final-state 
interactions and permits the extraction of free neutron structure \cite{Sargsian:2005rm}. 
In tagged DIS in fixed-target experiments, 
the proton emerges with typical momenta $|\bm{p}_p| \lesssim$ 100 MeV and is captured 
with special detectors placed close to the target (JLab 6/12 GeV 
BONuS \cite{Baillie:2011za,Tkachenko:2014byy,Bonus12}, ALERT \cite{Armstrong:2017zqr}).
In collider experiments at EIC, the proton moves forward with $\sim 1/2$ the
deuteron beam momentum and is detected with forward detectors integrated in the 
interaction region and beam optics. The collider setup offers many advantages for tagging
(no target material, acceptance at proton rest-frame momenta $|\bm{p}_p| \sim 0$, 
good momentum resolution), and the physics potential was studied in 
an R\&{D} project \cite{LD1506,Cosyn:2014zfa}. With the possibility of polarized deuteron beams at EIC,
it is interesting to assess the potential of polarized tagged DIS for precise measurements 
of neutron spin structure.

In this letter we report about the development of a theoretical framework for neutron spin structure 
measurements with polarized tagged DIS in collider experiments; see Ref.~\cite{Frankfurt:1983qs} for 
an earlier study. Using methods of light-front (LF) nuclear structure, we calculate the longitudinal 
spin asymmetries in polarized tagged DIS and study the dependence on the measured proton momentum. 
We consider the asymmetries formed with all three deuteron spin states ($\pm 1, 0$) and the 
two maximum-spin states only ($\pm 1$). We show that the proton momentum can be used to
select pure S-wave configurations and eliminate D-wave depolarization. We discuss how
free neutron spin structure can be extracted through pole extrapolation of the asymmetries.
Details will be provided in a forthcoming article \cite{CW}.
\section{Spin asymmetries in polarized tagged DIS}
\label{sec:spin_asymmetries}
\begin{figure}[t]
\begin{tabular}{ll}
\parbox[c]{.12\textwidth}{\includegraphics[width=.12\textwidth]{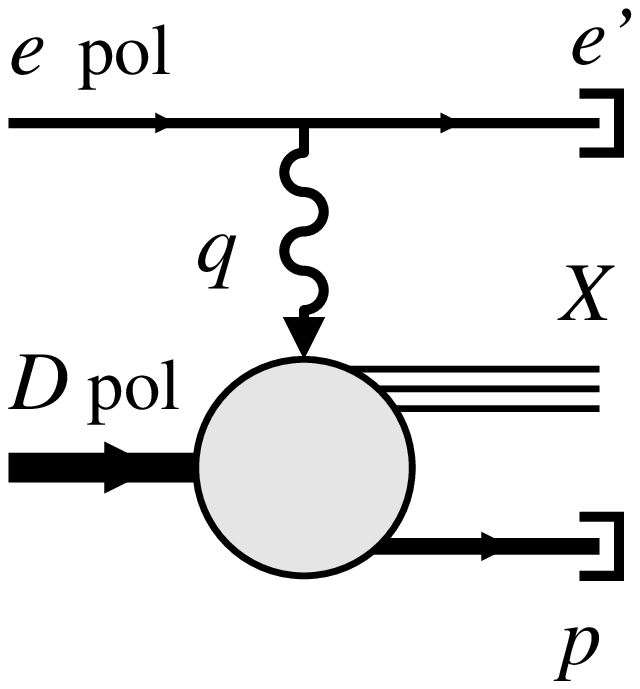}} 
\hspace{.01\textwidth} 
&
\parbox[c]{.3\textwidth}{\includegraphics[width=.3\textwidth]{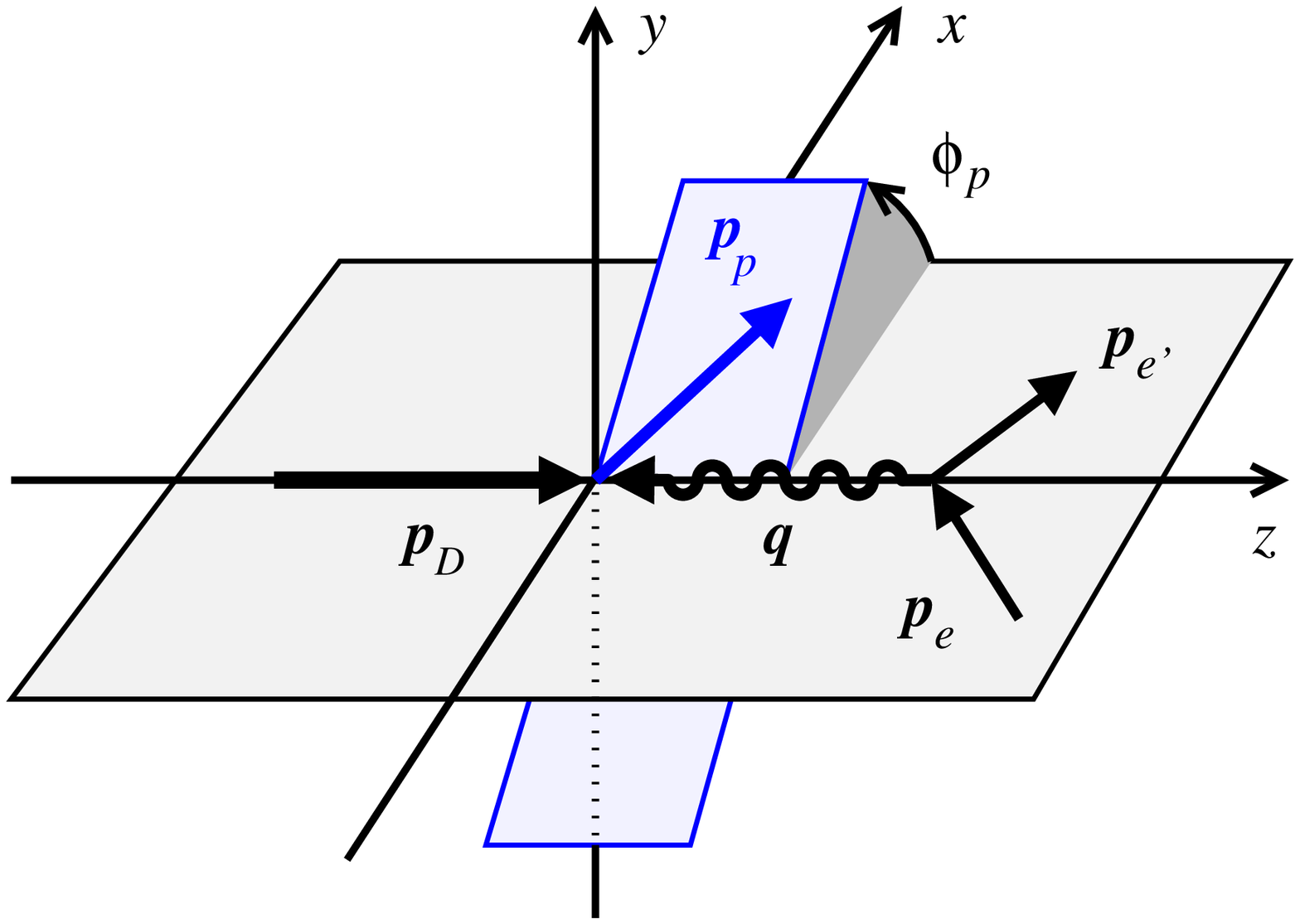}}
\\[-2ex]
{\footnotesize (a)} 
&
{\footnotesize (b)} 
\end{tabular}
\caption{\label{fig:electron_deuteron_spins}
(a) Polarized electron-deuteron DIS with proton tagging, $e + D \rightarrow e' + X + p$.
(b) Momenta in the photon-deuteron collinear frame.}
\end{figure}

Polarized tagged DIS with deuteron 4-momentum $p_D$ and 4-momentum 
transfer $q \equiv p_e - p_e'$ is described by the invariant differential cross section
(see Fig.~\ref{fig:electron_deuteron_spins}a)
\begin{align}
\text{d}\sigma [eD \rightarrow e'Xp] &= {\mathcal F} \text{d}x \, \text{d}Q^2 \, 
\text{d}\Gamma_p, \hspace{2em}
\label{cross_section}
\\[1ex]
{\mathcal F} &\equiv {\mathcal F}(x, Q^2; \{ p_p \}; \textrm{pol}_e, \textrm{pol}_D),
\end{align}
where $x \equiv Q^2/(p_D q)$ and $Q^2 \equiv -q^2$ 
are the usual DIS variables, $\{ p_p \}$ denotes a set of variables describing
the measured proton momentum, $\text{d}\Gamma_p$ is the invariant phase space element in the proton momentum, 
and ``$\textrm{pol}_e$'' and ``$\textrm{pol}_D$'' denote generic variables specifying 
the initial electron and deuteron polarization. As generally for 
semi-inclusive DIS, it is convenient to describe the process in a frame where the 3-vectors
$\bm{p}_D$ and $\bm{q}$ are collinear along the $z$-axis (see Fig.~\ref{fig:electron_deuteron_spins}b). 
We specify the particle momenta 
in this frame by their LF components $p^\pm \equiv p^0 \pm p^3, \, \bm{p_T} \equiv (p^1, p^2)$.
The proton is described by its plus momentum fraction and transverse momentum
\begin{align}
\alpha_p \; \equiv \; 2 p_p^+ / p_D^+  
\hspace{1em}
(0 < \alpha_p < 2), \hspace{1em}
\bm{p}_{pT} ,
\end{align}
with $\text{d}\Gamma_p = (\text{d}\alpha_p / \alpha_p) \, \text{d}^2 p_{pT}$.
The typical ranges of these variables are $|\alpha_p - 1| \lesssim 0.1$ and $|\bm{p}_{pT}| \lesssim$ 100 MeV,
corresponding to proton 3-momenta $|\bm{p}_p| \lesssim$ 100 MeV in the deuteron rest frame.
The dependence of the cross section on the proton azimuthal angle $\phi_p$ is kinematic and can be established 
on general grounds; the decomposition will be presented elsewhere \cite{CW}. In the following we
consider the azimuthally integrated cross section, 
\begin{align}
\text{d}\sigma \; \equiv \; \left[ {\textstyle \int} \text{d}\phi_p \mathcal{F} \right] \; \text{d}x \, \text{d}Q^2 
\; (\text{d}\alpha_p/\alpha_p )\; 
|\bm{p}_{pT}| \; \text{d}|\bm{p}_{pT}| ,
\label{phi_integrated}
\end{align}
which is sufficient for neutron spin structure measurements.
%
%

The polarization variables in Eq.~(\ref{cross_section}) have to be specified depending on the experimental setup.
We consider a collider experiment in which the electron and deuteron beams move in opposite directions 
along the same axis in the laboratory (zero crossing angle). The electron entering the scattering
process is in a spin state quantized along the beam axis with projections $\Lambda_e = \pm \frac{1}{2}$. 
The deuteron is in a spin state quantized along the beam axis with projections $\Lambda_D = (1, 0, -1)$;
combinations of these pure states correspond to spin ensembles with longitudinal vector and tensor polarization. 
We denote the cross section in the pure spin states by
\begin{align}
\text{d}\sigma \equiv \text{d}\sigma (\Lambda_e, \Lambda_D) .
\end{align}
The beam polarizations in the laboratory frame determine the effective polarization of the electron and 
deuteron in the photon-deuteron collinear frame; the corresponding kinematic factors (depolarization factors)
are given below. 

Experiments measure differences and sums of the cross sections in different electron and deuteron 
polarization states and their ratios (spin asymmetries). The ``polarized'' cross section is calculated 
as the difference between the $\Lambda_D = \pm 1$ deuteron spin states,\footnote{In fully inclusive DIS 
the spin dependence of the cross section is entirely through double-spin dependent terms 
$\propto \Lambda_e \Lambda_D$, and it is sufficient to take the difference in the electron or deuteron 
spin alone, with the other spin remaining fixed. In tagged DIS the cross section can have 
also single-spin dependent terms $\propto \Lambda_e$ and $\propto \Lambda_D$, 
and it is necessary to take the double difference 
in the electron and deuteron spins in order to isolate the double-spin dependent terms.}
\begin{align}
\hspace{-.5em}
\text{d}\sigma_\parallel \equiv 
{\textstyle\frac{1}{4}} [ 
  \text{d}\sigma ({\textstyle\frac{1}{2}}, 1)
- \text{d}\sigma (-{\textstyle\frac{1}{2}}, 1)
- \text{d}\sigma ({\textstyle\frac{1}{2}}, -1)
+ \text{d}\sigma (-{\textstyle\frac{1}{2}}, -1) ] .
\label{sigma_polarized}
\end{align}
``Unpolarized'' cross sections can be formed in two ways: (i)~as the sum of all three deuteron spin
states ($\Lambda_D = \pm 1, 0$), 
\begin{align}
\text{d}\sigma_3 &\equiv \textstyle{\frac{1}{6}} \sum\limits_{\Lambda_e = \pm 1/2}
[\text{d}\sigma (\Lambda_e, +1) + \text{d}\sigma (\Lambda_e, -1) + \text{d}\sigma (\Lambda_e, 0)];
\label{sigma_3}
\end{align}
(ii)~as the sum of the two maximum-spin states only ($\Lambda_D = \pm 1$),
which enter in the polarized cross section Eq.~(\ref{sigma_polarized}),
\begin{align}
\text{d}\sigma_2 &\equiv \textstyle{\frac{1}{4}}
\sum\limits_{\Lambda_e = \pm 1/2}
[\text{d}\sigma (\Lambda_e, +1) + \text{d}\sigma (\Lambda_e, -1)] .
\label{sigma_2}
\end{align}
The combination Eq.~(\ref{sigma_2}) implies a nonzero tensor polarization of the deuteron ensemble.
Correspondingly, one can define the ``three-state'' and ``two-state'' tagged spin asymmetries as
\begin{align}
A_{\parallel, 3} \equiv \frac{\text{d}\sigma_\parallel}{\text{d}\sigma_{3}}, 
\hspace{2em}
A_{\parallel, 2} \equiv \frac{\text{d}\sigma_\parallel}{\text{d}\sigma_{2}}. 
\label{asymmetries}
\end{align}
They depend on $x$ and $Q^2$ and the proton momentum variables $\alpha_p$
and $|\bm{p}_{pT}|$ [$\phi_p$ is integrated out, Eq.~(\ref{phi_integrated})],
\begin{align} 
A_{\parallel, i}(x, Q^2; \alpha_p, |\bm{p}_{pT}|) \hspace{2em} (i = 3, 2).
\end{align} 
In the following we compute the tagged asymmetries in an approach that separates
nuclear and nucleonic structure, study their dependence on the proton momentum, 
and assess their usefulness for neutron spin structure extraction.
\section{Deuteron light-front wave function}
\begin{figure}[t]
\includegraphics[width=.48\textwidth]{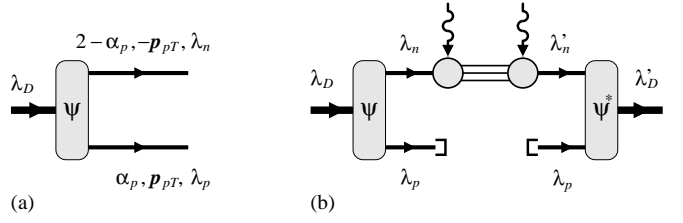}
\caption{\label{fig:wf_ia}
(a) Deuteron LF wave function. (b) Polarized tagged DIS cross section in impulse approximation.}
\end{figure}

To calculate the tagged DIS cross section we use methods of LF nuclear structure.
The quantization scheme is unique in that the effects of energy nonconservation (or 4-momentum nonconservation) 
in intermediate states remain finite in the high-energy limit and enables a practical description 
of high-energy scattering from composite systems \cite{Frankfurt:1983qs}. The technique is summarized 
in Ref.~\cite{Strikman:2017koc}; here we only describe the aspects specific to the spin degrees of freedom.
The calculation is performed in the photon-deuteron collinear frame (see Fig.~\ref{fig:electron_deuteron_spins}b). 
The deuteron is described by a wave function in nucleonic degrees of freedom at fixed LF time
(see Fig.~\ref{fig:wf_ia}a). The nucleon spin states are chosen as LF helicity states 
(LF boosts of rest-frame spin states quantized along the $z$-axis \cite{Soper:1972xc,Brodsky:1997de}) 
with helicity quantum numbers $\lambda_p, \lambda_n = \pm \frac{1}{2}$; the deuteron LF helicity
$\lambda_D = (1, 0, -1)$ is identical to its ordinary spin projection on the $z$-axis (its transverse
momentum is zero). The LF wave function is constructed from a rotationally covariant
3-dimensional wave function in the center-of-mass (CM) frame of the $pn$ pair \cite{Kondratyuk:1983kq} 
\begin{align}
\Psi (\alpha_p , \bm{p}_{pT}; \lambda_p, \lambda_n | \lambda_D)
&= \sum_{\mu_p, \, \mu_n} 
\widetilde{\Psi} (\bm{k}, \mu_p, \mu_n | \lambda_D ) 
\nonumber \\
& \times U^\ast (\bm{k}, \mu_p, \lambda_p ) \; U^\ast(-\bm{k}, \mu_n, \lambda_n),
\label{wf_k_3d_lf}
\end{align}
\vspace{-3ex}
\begin{align}
\widetilde{\Psi} (\bm{k}, \mu_p, \mu_n | \lambda_D ) 
& \equiv \frac{\epsilon^a (\lambda_D)}{\sqrt{2}} \left[ \delta^{ab} f_0(k) 
+ \left( \frac{3 k^a k^b}{|\bm{k}|^2} - \delta^{ab} \right) \frac{f_2(k)}{\sqrt{2}} \right] 
\nonumber \\
& \times \; \chi^\dagger (\mu_n) \left[ \sigma^b (i \sigma^2) \right] \chi^\ast (\mu_p) ,
\label{wf_k_3d}
\end{align}
\vspace{-3ex}
\begin{align}
U (\bm{k}, \mu_p, \lambda_p ) & \equiv \chi^\dagger (\mu_p) \left[
\frac{E + k^3 + m + \bm{k}_T \bm{\sigma}_T \sigma^3}{\sqrt{2 (E + k^3) (E + m)}} \right] \chi(\lambda_p) ,
\label{U_proton}
\\
U (-\bm{k}, \mu_n, \lambda_n ) & \equiv \; [\textrm{same with $\bm{k} \rightarrow -\bm{k}$; \,
$\mu_p, \lambda_p \rightarrow \mu_n, \lambda_n$}] .
\label{U_neutron}
\end{align}
Here $\bm{k} = (\bm{k}_T, k^3)$ is the CM momentum of the $pn$ pair and related to the LF momentum 
variables by ($m$ is nucleon mass)
\begin{align}
\alpha_p = 1 + k^3/E, \hspace{2em} E = \sqrt{m^2+|\bm{k}|^2}, \hspace{2em} \bm{p}_{pT} = \bm{k}_T .
\end{align}
$\widetilde{\Psi}$ is the rotationally covariant wave function in the CM frame;
it is formulated in canonical nucleon spin states with quantum numbers $\mu_p, \mu_n = \pm \frac{1}{2}$.
$\sigma^a \; (a = 1, 2, 3)$ are the Pauli spin matrices; $\chi (\nu) \; (\nu = \mu_p, \mu_n, \lambda_p, \lambda_n)$ 
are the two-component spinors for spin projection $\nu = \pm \frac{1}{2}$ on the $z$-axis,
$\chi(\textstyle{\frac{1}{2}}) = (1, 0)^T, \; \chi(\textstyle{-\frac{1}{2}}) = (0, 1)^T$;
$\bm{\epsilon} (\lambda_D)$ is the spin-1 polarization vector for spin projection $\lambda_D$,
$\bm{\epsilon} (\pm 1) = {\textstyle\frac{1}{\sqrt{2}}}(\mp 1, -i, 0)^T, \,
\bm{\epsilon} (0) = (0, 0, 1)^T$.
The CM frame wave function Eq.~(\ref{wf_k_3d}) involves the S- and D-waves of the $pn$ pair.
They are described by the radial wave functions $f_0(k)$ and $f_2(k)$, $k \equiv |\bm{k}|$, 
which are normalized as
\begin{align}
4\pi \int_0^\infty \frac{\text{d}k \, k^2}{E(k)} [f_0^2(k) + f_2^2(k)] \;\; = \;\; 1 .
\label{wf_k_3d_normalization}
\end{align}
In Eqs.~(\ref{wf_k_3d_lf})--(\ref{U_neutron}) $U$ are the momentum-dependent Melosh rotations connecting 
the canonical nucleon spin states with the LF helicity states ($U^\ast$ denotes the complex conjugate). 
The construction Eqs.~(\ref{wf_k_3d_lf})--(\ref{U_proton}) effectively implements rotational 
invariance in the LF calculation and permits a simple description of
the deuteron spin structure in this context \cite{Frankfurt:1981mk,Kondratyuk:1983kq,Keister:1991sb}. 
%
%

The radial wave functions can be obtained by solving the 2-body bound-state equation with $pn$ interactions
at fixed LF time \cite{Frankfurt:1981mk,Cooke:2001kz}. 
Alternatively, one can approximate them by the nonrelativistic wave functions as
\begin{align}
f_L(k) \;\; \stackrel{\rm app.}{=} \;\; \sqrt{E(k)} \, f_{L, \, \textrm{nr}}(k) \hspace{2em} (L = 0, 2);
\label{nonrel_approx}
\end{align}
the factor arises from the normalization Eq.~(\ref{wf_k_3d_normalization}). Equation~(\ref{nonrel_approx})
is reliable at all momenta of interest here \cite{Frankfurt:1981mk,Strikman:2017koc};
in the numerical studies we use it with the AV18 wave functions \cite{Wiringa:1994wb}.
\section{Deuteron light-front spectral function}
\label{sec:spectral}
We calculate the tagged DIS cross section in the impulse approximation, in which the DIS final 
state and the spectator evolve independently after the scattering process (see Fig.~\ref{fig:wf_ia}b). 
The general expressions for the cross section and scattering tensor are summarized in Ref.~\cite{Strikman:2017koc}. 
The polarized deuteron scattering tensor is factorized in the polarized neutron scattering tensor 
and the deuteron LF spectral function,
\begin{align}
& \mathcal{S} (\alpha_p , \bm{p}_{pT}; \lambdanp, \lambda_n | 
\lambdaDp, \lambda_D)
\; \equiv \; \frac{1}{2 - \alpha_p} 
\nonumber \\
&\times \sum_{\lambda_p} 
\Psi^\ast (\alpha_p , \bm{p}_{pT}; \lambda_p, \lambdanp | \lambdaDp) \, 
\Psi (\alpha_p , \bm{p}_{pT}; \lambda_p, \lambda_n | \lambda_D) .
\label{spectral_general}
\end{align}
It describes the conditional probability density of the neutron in the deuteron when removing 
a proton with LF momentum $\alpha_p$ and $\bm{p}_{pT}$, and is normalized as
\begin{align}
\sum_{\lambda_n} 
\int\limits_0^2\frac{\text{d}\alpha_p}{\alpha_p} \int \text{d}^2p_{pT} \,
\mathcal{S} (\alpha_p , \bm{p}_{pT}; \lambda_n, \lambda_n | \lambdaDp, \lambda_D) 
\, = \, \delta_{\lambdaDp\lambda_D}.
\label{normalization}
\end{align}

The general spectral function Eq.~(\ref{spectral_general}) is a matrix in the LF helicities of the external 
deuteron and intermediate neutron states. In the calculation of the cross section, the deuteron LF helicities 
are averaged over with the deuteron spin density matrix, which describes the effective deuteron polarization
in LF helicities, taking into account that the experimental polarization is along the beam 
axis (see Sec.~\ref{sec:spin_asymmetries}). The neutron helicities are averaged over with the spin structure 
of the neutron scattering tensor. In the particular case of the $\phi_p$-integrated cross 
sections with the longitudinal deuteron polarization described in Sec.~\ref{sec:spin_asymmetries}, 
the result can be expressed in terms of three distinct LF helicity projections of the spectral function
Eq.~(\ref{spectral_general}) \cite{CW}:
\begin{align}
\mathcal{S}_U &= \sum_{\lambda_D\lambdaDp} (\rho_U)_{\lambda_D\lambdaDp} \sum_{\lambda_n\lambdanp}
\mathcal{S} (\lambdanp, \lambda_n | \lambdaDp, \lambda_D) \; \delta_{\lambda_n\lambdanp} ,
\label{S_U_def}
\\
\Delta\mathcal{S}_S &= \sum_{\lambda_D\lambdaDp} (\rho_S)_{\lambda_D\lambdaDp} \sum_{\lambda_n\lambdanp}
\mathcal{S} (\lambdanp, \lambda_n | \lambdaDp, \lambda_D) 
\, (2 \lambda_n) \, \delta_{\lambda_n\lambdanp} ,
\label{Delta_S_S_def}
\\
\mathcal{S}_T &= \sum_{\lambda_D\lambdaDp} (\rho_T)_{\lambda_D\lambdaDp}
\sum_{\lambda_n\lambdanp}
\mathcal{S} (\lambdanp, \lambda_n | \lambdaDp, \lambda_D) \; \delta_{\lambda_n\lambdanp} .
\label{S_T_def}
\end{align}
Here $\rho_{U, \, S, \, T}$ are the spin-1 density matrices describing an unpolarized, 
vector-polarized, and tensor-polarized ensemble quantized along the $z$-axis,
\begin{align}
(\rho_U)_{\lambda_D\lambdaDp} &= {\textstyle\frac{1}{3}} \textrm{diag}(1,1,1),
\\
(\rho_S)_{\lambda_D\lambdaDp} &= {\textstyle\frac{1}{2}} \textrm{diag}(1,0,-1),
\\
(\rho_T)_{\lambda_D\lambdaDp} &= {\textstyle\frac{1}{6}} \textrm{diag}(1,-2,1).
\end{align}
The functions $\mathcal{S}_U, \Delta \mathcal{S}_S$ and $\mathcal{S}_T$ depend on 
the proton LF momentum variables $\alpha_p$ and $|\bm{p}_{pT}|$. $\mathcal{S}_U$ describes 
the LF helicity-independent probability of neutrons in an unpolarized deuteron ensemble; 
$\mathcal{S}_T$ describes the LF helicity-independent probability of neutrons in a 
tensor-polarized deuteron ensemble; $\Delta\mathcal{S}_S$ describes the LF helicity-dependent 
probability of neutrons in a vector-polarized deuteron ensemble. Their interpretation
is similar to that of the unpolarized and helicity-polarized parton densities in the 
nucleon and can be developed further along these lines \cite{CW}.
We evaluate the functions by using the explicit expression of the deuteron LF wave function 
and performing the sums over the LF helicities and obtain
\cite{CW}
\begin{align}
\mathcal{S}_U (\alpha_p, |\bm{p}_{pT}|)  \; &= \; \frac{1}{2-\alpha_p} \left(f_0^2 + f_2^2 \right),
\label{P_unpol}
\\
\Delta \mathcal{S}_S (\alpha_p, |\bm{p}_{pT}|)  
\; &= \; \frac{1}{2-\alpha_p}\left( f_0 - \frac{f_2}{\sqrt{2}}\right) \left( C_0 f_0 
- \frac{C_2 f_2}{\sqrt{2}} \right), 
\label{P_vector}
\\
C_0
\; &= \; 1 - \frac{(E + k^3) |\bm{k}_T|^2}{(E + m)(m^2 + |\bm{k}_T|^2)},
\label{C_0}
\\
C_2
\; &= \; 1 - \frac{(E + 2 m)(E + k^3) |\bm{k}_T|^2}{(m^2 + |\bm{k}_T|^2) |\bm{k}|^2},	
\label{C_2}
\\
\mathcal{S}_T (\alpha_p, |\bm{p}_{pT}|)
\; &= \; \frac{1}{2-\alpha_p} C_T \left( -2 f_0 - \frac{f_2}{\sqrt{2}} \right) \frac{f_2}{\sqrt{2}},
\label{P_tensor}
\\
C_T &= 1 - \frac{3 |\bm{k}_T|^2}{2 |\bm{k}|^2} ,
\label{C_T}
\end{align}
where $f_{0, 2} \equiv f_{0, 2}(k)$ are the radial wave functions of Eq.~(\ref{wf_k_3d}).
The factors $C_0$ and $C_2$ in Eq.~(\ref{P_vector}), and $C_T$ in Eq.~(\ref{P_tensor}), arise from the
contraction of the Melosh rotations Eq.~(\ref{U_proton}) and contain the relativistic spin 
effects in the neutron distributions in the polarized deuteron. They play an essential role in the
proton momentum dependence of the polarized tagged DIS cross section (see below).

The deuteron spectral function Eq.~(\ref{P_unpol}) satisfies sum rules for the total
baryon number and the total LF momentum of the neutron, obtained by integrating over the
proton momentum,
\begin{align}
\int\frac{d\alpha_p}{\alpha_p}\int d^2 p_{pT} \, \mathcal{S}_U (\alpha_p, \bm{p}_{pT}) \times
\left\{ \! \begin{array}{c} 1 \\[0ex] (2 - \alpha_p) \end{array} \! \right\} \, = \, 
\left\{ \! \begin{array}{r} 1 \\[0ex] 1 \end{array} \! \right\} .
\label{sum_rules}
\end{align}
They follow from the normalization condition Eq.~(\ref{normalization}) and the symmetry of the
deuteron wave function under $\alpha_p \rightarrow 2 - \alpha_p$, which is a consequence of 
the rotational invariance encoded in Eq.(\ref{wf_k_3d}). The sum rules indicate that a 
consistent description of deuteron structure in terms of nucleon degrees of freedom is achieved 
in the impulse approximation.
\section{Deuteron structure in spin asymmetries}
\label{sec:deuteron_structure}
Using the deuteron spectral functions we compute the longitudinal spin asymmetries of the
azimuthally integrated tagged DIS cross section (see Sec.~\ref{sec:spin_asymmetries}).
We consider the DIS limit $Q^2 \rightarrow \infty$, $x$ fixed, and neglect power-suppressed terms
in the kinematic factors and the spin-dependent cross section ($g_2$ structure function).
The result for the three- and two-state asymmetries Eq.~(\ref{asymmetries}) can be expressed 
in simple form as [here $i = 3, 2$]
\begin{align}
A_{\parallel, i}(x, Q^2; \alpha_p, |\bm{p}_{pT}|) 
\; =& \; A_{\parallel n}(\tilde{x}, Q^2) \; \mathcal{D}_i(\alpha_p, |\bm{p}_{pT}|) , 
\label{asymmetry_factorized}
\\
A_{\parallel n}(\tilde{x}, Q^2) \; =& \;
\frac{D_\parallel \, g_{1n}(\tilde{x},Q^2)}{2(1+\epsilon R_n)F_{1n}(\tilde{x},Q^2)} .  
\label{A_n}
\end{align}
$A_{\parallel n}$ is the longitudinal spin asymmetry for DIS on the free neutron. It is given 
in terms of the polarized and unpolarized neutron structure functions, $g_{1n}$ and $F_{1n}$, 
and the neutron L/T ratio $R_n$. The neutron functions are evaluated at the modified $x$-value
\begin{align}
\tilde{x}=x/(2-\alpha_p),
\end{align}
where $2 - \alpha_p$ is the plus momentum fraction of the active neutron, whose value is 
fixed by the $\alpha_p$ of the tagged proton. In Eq.~(\ref{A_n})
$\epsilon$ is the virtual photon polarization parameter; $D_\parallel$ is the depolarization
factor describing the effective polarization in the photon-deuteron collinear frame induced 
by the experimental polarization along the beam axis (see Sec.~\ref{sec:spin_asymmetries}),
\begin{align}
\epsilon = \frac{1 - y}{1 - y + y^2/2},
\hspace{2em}
D_\parallel = \frac{2y(1-y/2)}{1-y+y^2/2} ;
\end{align}
the scaling variable $y \equiv (p_D q)/(p_D p_e)$ is the same for DIS on the deuteron and the 
free neutron up to power-suppressed corrections (we do not distinguish between the two in the notation). 
The typical magnitude of $A_{\parallel n}$ is a few $\times 10^{-2}$ for 
$\tilde{x} \gtrsim 10^{-2}$ and $y = \mathcal O(1)$ \cite{Cosyn:2014zfa}. 
The deuteron structure effects in Eq.~(\ref{asymmetry_factorized})
are contained in the factors $\mathcal{D}_i \; (i = 3,2)$, which depend on the proton variables
$\alpha_p$ and $|\bm{p}_{pT}|$. They play the role of a ``dynamical depolarization factor'' specific 
to the deuteron configuration selected by the tagged proton momentum. For the three-state 
and two-state asymmetries, Eq.~(\ref{asymmetries}), they are obtained as
\begin{align}
\mathcal{D}_3 (\alpha_p, |\bm{p}_{pT}|)
\; &\equiv \;
\frac{\Delta\mathcal{S}_S (\alpha_p, |\bm{p}_{pT}|)}{\mathcal{S}_U (\alpha_p, |\bm{p}_{pT}|)}\,,
\label{D_3}
\\
\mathcal{D}_2 (\alpha_p, |\bm{p}_{pT}|)
\; &\equiv \;
\frac{\Delta\mathcal{S}_S (\alpha_p, |\bm{p}_{pT}|)}{[\mathcal{S}_U
+ \mathcal{S}_T] (\alpha_p, |\bm{p}_{pT}|)} .
\label{D_2}
\end{align}
The numerator in both cases is the vector-polarized spectral function Eq.~(\ref{Delta_S_S_def}). 
The denominator in the three-state asymmetry is the unpolarized spectral function Eq.~(\ref{S_U_def});
in the two-state asymmetry there is a contribution of the tensor-polarized spectral function
Eq.~(\ref{S_T_def}), because the sum of $\pm 1$ spin states only (without the 0 state) corresponds 
to a spin ensemble with nonzero tensor polarization. 

%
%
\begin{figure}[t]
\begin{center}
\begin{tabular}{l}
\includegraphics[width=.4\textwidth]{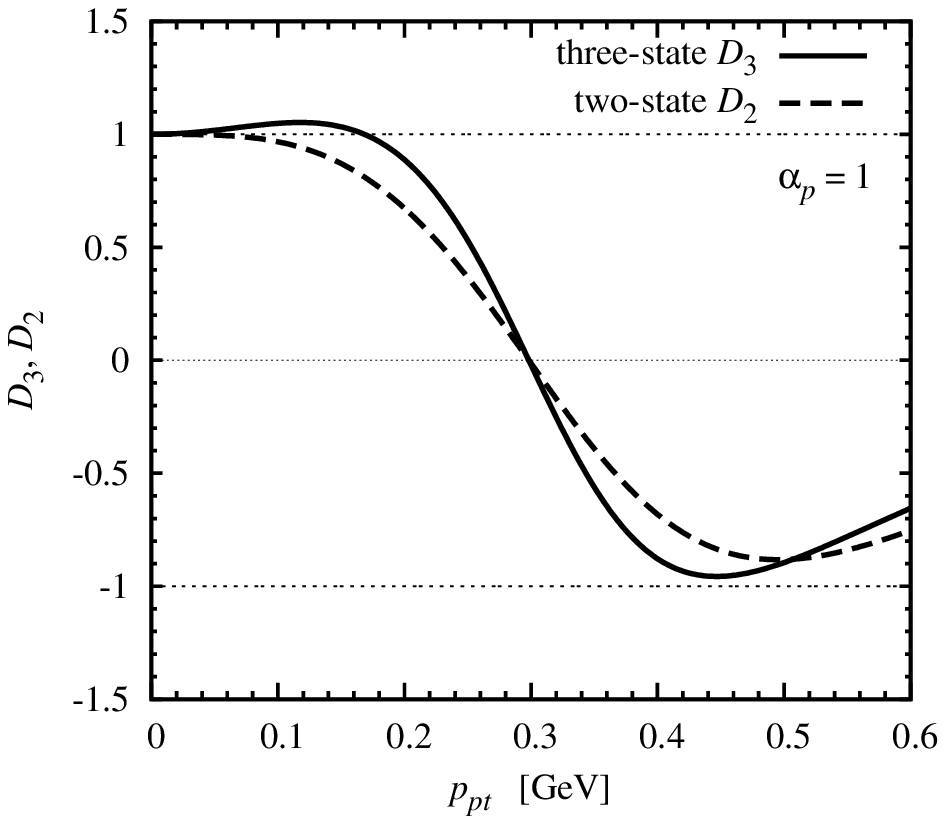} 
\\[-5ex] 
{\footnotesize (a)}
\\[3ex]
\includegraphics[width=.4\textwidth]{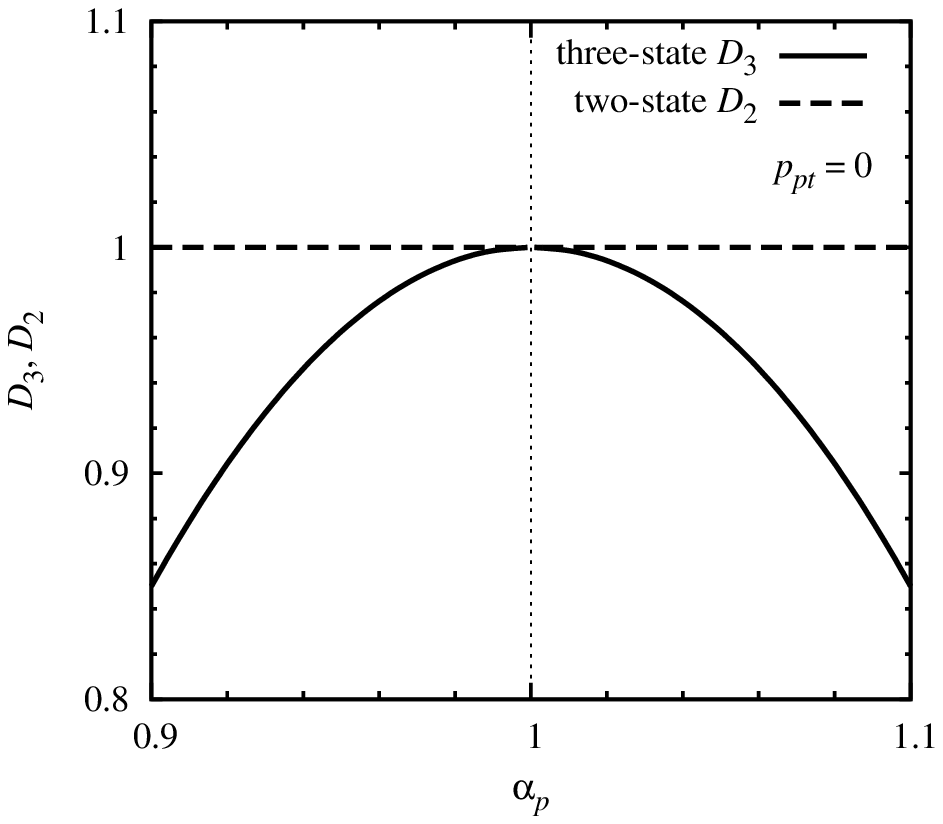}
\\[-5ex] 
{\footnotesize (b)}
\end{tabular}
\end{center}
\caption{Deuteron structure factors $\mathcal{D}_3$ (three-state asymmetry) and $\mathcal{D}_2$ (two-state asymmetry)
in polarized tagged DIS, Eqs.~(\ref{D_3}) and (\ref{D_2}). (a) As functions of $p_{pT} \equiv |\bm{p}_{pT}|$,
for fixed $\alpha_p = 1$. (b) As functions of $\alpha_p$ in the vicinity of $\alpha_p = 1$, 
for fixed $p_{pT} = 0$.}
\label{fig1}
\end{figure}
We evaluate the deuteron structure factors Eqs.~(\ref{D_3}) and (\ref{D_2}) using the explicit 
expressions for the spectral functions Eqs.~(\ref{P_unpol})--(\ref{C_T}). Important differences
between $\mathcal{D}_3$ and $\mathcal{D}_2$ can be deduced from the analytic expressions. 
(a)~$\mathcal{D}_2$ is bounded by unity,
\begin{equation}
-1 \leq \mathcal{D}_2 \leq 1.
\label{eq:bound}
\end{equation}
No such bound is found for $\mathcal{D}_3$, which attains absolute values larger than unity.
(b) ${\mathcal D}_2$ is equal to unity at zero proton transverse momentum and arbitrary plus momentum,
\begin{align}
\mathcal{D}_2 (\alpha_p, |\bm{p}_{pT}| = 0) \; = \; 1
\hspace{2em} \textrm{($\alpha_p$ arbitrary).}
\label{D_2_pt0}
\end{align}
This happens because the factors $C_0, C_2$ and $C_T$ in Eqs.~(\ref{C_0}), (\ref{C_2}) and (\ref{C_T})
become unity at $\bm{k}_T = 0$ [the Melosh rotations are trivial at zero transverse momentum,
$U(\bm{k}_T = 0) = 1$, cf.\ Eqs.~(\ref{U_proton}) and (\ref{U_neutron})]. 
In contrast, $\mathcal{D}_3$ is equal to unity only at $|\bm{p}_{pT}| = 0$ and $\alpha_p = 1$.
(c) At small proton momenta, $|\alpha_p - 1| \ll 1$ and $|\bm{p}_{pT}| \ll m$, corresponding to
$|\bm{k}| \ll m$, the D-wave affects $\mathcal{D}_2$ only at quadratic order, but $\mathcal{D}_3$ 
already at linear order,
\begin{align}
\left.
\begin{array}{lcl}
\mathcal{D}_2 \; &=& \; 1 \; + \; \textrm{terms} \; f_2^2/f_0^2
\\[1ex]
\mathcal{D}_3 \; &=& \; 1 \; + \; \textrm{terms} \; f_2/f_0
\end{array}
\right\} \hspace{2em} (|\bm{k}| \ll m).
\end{align}
This can be demonstrated by expanding the factors $C_0$ and $C_2$ in Eqs.~(\ref{C_0}) and (\ref{C_2})
in $|\bm{k}|/m$, and then expanding the spectral function ratios in Eqs.~(\ref{D_3}) and (\ref{D_2}) 
in $f_2/f_0$.

Figure~\ref{fig1}a shows the deuteron structure factors Eqs.~(\ref{D_3}) and (\ref{D_2})
as functions of $|\bm{p}_{pT}|$ for fixed $\alpha_p = 1$. 
One observes: (a)~Both $\mathcal{D}_3$ and $\mathcal{D}_2$ are unity at 
at $\alpha_p = 1$ and $|\bm{p}_{pT}| = 0$, where $|\bm{k}| = 0$ and only the S-wave is present. 
(b) $\mathcal{D}_3$ and $\mathcal{D}_2$ 
remain close to unity for $|\bm{p}_{pT}| \lesssim$ 150 MeV, where the S-wave dominates. The D-wave 
contribution raises $\mathcal{D}_3$ above unity but lowers $\mathcal{D}_2$, in accordance with the 
bound Eq.~(\ref{eq:bound}), showing the effect of the tensor polarized structure in $\mathcal{D}_2$.
(c) Both $\mathcal{D}_3$ and $\mathcal{D}_2$ decrease significantly at $|\bm{p}_{pT}| \gtrsim$ 150 MeV and
pass through zero at $|\bm{p}_{pT}| \approx$ 300 MeV, where the combination $(f_0 - f_2/\sqrt{2})$ vanishes.
Both become negative at larger momenta, where the D-wave dominates.

Figure~\ref{fig1}b shows the deuteron structure factors as functions of $\alpha_p$ near $\alpha_p = 1$
for fixed $|\bm{p}_{pT}| = 0$. One observes that $\mathcal{D}_2 = 1$ in accordance with
Eq.~(\ref{D_2_pt0}), whereas $\mathcal{D}_3$ shows a sizable variation in $\alpha_p - 1$.
Altogether we find that at small proton momenta $\mathcal{D}_2$ is much closer to unity 
than $\mathcal{D}_3$.

In the present analysis we consider the longitudinal spin asymmetries $A_{\parallel, i} (i = 2, 3)$ 
in the DIS limit $Q^2 \gg m^2$ and neglect power corrections (leading-twist approximation). 
In this approximation the $A_{\parallel, i}$ only involve the $g_{1n}$ structure function;
the contribution from $g_{2n}$ is power-suppressed \cite{Anselmino:1994gn}; and the variables
characterizing the DIS process on the neutron are taken at their $Q^2 \gg m^2$ values.
At this level of accuracy the effects of interactions in the transverse component of the electromagnetic 
current operator in LF quantization can generally be neglected, enabling a consistent description of 
deuteron structure with nucleon degrees of freedom only \cite{Frankfurt:1988nt}; see also Ref.~\cite{Lev:1998qz}. 
This is seen e.g.\ in the fact that the deuteron spectral function in the impulse approximation satisfies 
the LF momentum sum rule Eq.(\ref{sum_rules}), which in turn ensures the momentum sum rule for the deuteron structure 
function $F_{2D}$, provided that the kinematic limits are taken at 
$Q^2 \gg m^2$ \cite{Frankfurt:1988nt,Strikman:2017koc}.
The present analysis could be extended to calculate the longitudinal spin asymmetries including 
power corrections and the contributions from $g_{2n}$, or to calculate the transverse spin asymmetries,
which are altogether power-suppressed. At this level of accuracy the interaction effects in the 
transverse current component could no longer be neglected, and new considerations would be needed
in order to construct a consistent description with nucleons only --- implementation of rotational 
invariance in LF calculations with fixed particle number, or the so-called angular conditions on 
the current operators or matrix elements. These problems have been studied extensively in the 
context of elastic scattering on the deuteron; 
see Refs.~\cite{Kondratyuk:1983kq,Keister:1993mg,Lev:1998qz,Gilman:2001yh,Bakker:2002aw} and references therein;
the methods could be adapted to the case of DIS beyond leading power accuracy.
\section{Neutron spin structure from pole extrapolation}
The tagged DIS cross section is generally affected by initial-state nuclear modifications of the structure 
functions and final-state interactions of the DIS products with the spectator. These effects can be described
by dynamical models and added to the impulse approximation \cite{Strikman:2017koc,Cosyn:2017ekf}. 
An alternative approach is to eliminate them using the analytic properties of the proton 
momentum dependence \cite{Sargsian:2005rm}. Analytic continuation in the proton momentum can select 
$pn$ configurations at asymptotically large distances, where the neutron is effectively free. 
This approach does not require input beyond the impulse approximation and permits model-independent 
extraction of free neutron structure from tagged DIS. Its feasibility in the unpolarized case was studied 
in Refs.~\cite{Sargsian:2005rm,Strikman:2017koc}; here we apply it to the polarized case.

For the study of analytic properties one regards the deuteron wave function as a function of the 
invariant mass of the $pn$ pair,
\begin{align}
M_{pn}^2 \; &= \; \frac{4 (|\bm{p}_{pT}|^2 + m^2)}{\alpha_p (2 - \alpha_p)}
\; = \; 4 (|\bm{k}|^2 + m^2) ,
\label{inv_mass}
\end{align}
which depends on $\alpha_p$ and $|\bm{p}_{pT}|$ and takes values $M_{pn}^2 \geq 4 m^2$ for physical 
proton momenta. The wave function has a pole singularity of the type ($M_D$ is the deuteron mass;
we suppress the spin structure for the moment)
\begin{align}
\Psi (\alpha_p, |\bm{p}_{pT}|) \; &= \; \frac{\Phi}{M_{pn}^2 - M_D^2} \; + \; \textrm{[less singular]} ,
\label{pole_inv_mass}
\end{align}
which corresponds to $pn$ configurations of asymptotically large transverse spatial separation, 
outside the range of the nucleon interaction. The singularity lies outside the physical region of
proton momenta because $M_{pn}^2 \geq 4 m^2 > M_D^2$ but can be reached by analytic continuation in
$|\bm{p}_{pT}|^2$ to unphysical negative values,
\begin{align}
|\bm{p}_{pT}|^2 \; &\rightarrow \; -a_T^2, \hspace{2em}
a_T^2 \; \equiv \; m^2 - \alpha_p (2 - \alpha_p) M_D^2/4 .
\label{pole_pt}
\end{align}
The minimum value of $a_T^2$ occurs at $\alpha_p = 1$ and is $a_T^2 = m^2 - M_D^2/4 = m \epsilon_D
+ \mathcal{O}(\epsilon_D^2) = a^2$, where $\epsilon_D \equiv 2 m - M_D$ is the 
deuteron binding energy and $a^2 \equiv m \epsilon_D$ is the inverse squared Bethe-Peierls radius 
of the deuteron. Because of the small value of the deuteron binding energy ($\epsilon_D =$ 2.2 MeV) 
the singularity is very close to the physical region and can be reached by extrapolation in $|\bm{p}_{pT}|^2$. 
This opens a practical way of accessing non-interacting large-size $pn$ configurations 
in the deuteron through analytic continuation.
In the representation of the deuteron LF wave function in terms of the 3-dimensional 
CM frame wave function, Eq.~(\ref{wf_k_3d}), the pole Eq.~(\ref{pole_inv_mass}) appears through 
the pole of the S-wave radial function
\begin{align}
f_0(k) \; &= \; \frac{\sqrt{m} \, \Gamma}{|\bm{k}|^2 + a^2} \; + \; \textrm{[less singular]} ,
\label{pole_radial}
\end{align}
which is a general feature of the weakly bound system. This illustrates the close correspondence 
between the LF and the nonrelativistic description of the two-body bound state.

On general grounds, the tagged DIS cross section as a function of the proton momentum has a pole
$\sim (M_{pn}^2 - M_D^2)^{-2}$, corresponding to scattering from a large-size $pn$ pair.
The residue of this pole is, up to a constant factor, given by the {\em free neutron cross section}.
The pole is contained entirely in the impulse approximation to the tagged DIS cross section and 
is a feature of the deuteron spectral function (see Sec.~\ref{sec:spectral}). Final-state interactions 
do not modify the pole and affect only the less singular terms of the momentum dependence, because they 
involve momentum loop integrals \cite{Sargsian:2005rm,Strikman:2017koc}. To extract the free neutron 
structure functions, one measures the tagged DIS cross section for fixed $\alpha_p$ as a function of 
$|\bm{p}_{pT}|^2$ in the physical region $|\bm{p}_{pT}|^2 > 0$, removes the pole factor, and 
extrapolates the residue to $|\bm{p}_{pT}|^2 \rightarrow - a_T^2$ according to Eq.~(\ref{pole_pt})
(pole extrapolation). For unpolarized DIS the procedure was studied in detail 
in Refs.~\cite{Sargsian:2005rm,Strikman:2017koc}.

For polarized DIS the pole extrapolation can be performed at the level of the spin asymmetries. 
One measures the tagged spin asymmetries for fixed $\alpha_p$ as functions of $|\bm{p}_{pT}|^2$ 
for $|\bm{p}_{pT}|^2 > 0$. The pole factors $\sim (M_{pn}^2 - M_D^2)^{-2}$ in the cross sections 
cancel between the numerator and denominator of the spin asymmetries, 
as can be seen in the ratios of spectral functions, Eqs.~(\ref{D_3}) and (\ref{D_2}), 
so that the asymmetries are smooth functions in the limit $|\bm{p}_{pT}|^2 \rightarrow - a_T^2$.
The free neutron spin asymmetry is then obtained by extrapolating the measured tagged deuteron 
asymmetries to $|\bm{p}_{pT}|^2 = -a_T^2$ and removing the  deuteron structure factor.
Specifically, using the two-state deuteron asymmetry,
\begin{align}
& A_{\parallel n} (\tilde{x}, Q^2) \; = \; \frac{ 
A_{\parallel, 2}(x, Q^2; \alpha_p, 
|\bm{p}_{pT}|^2 \rightarrow - a_T^2 )}{\mathcal{D}_2 (\alpha_p, |\bm{p}_{pT}|^2 \rightarrow - a_T^2)} ,
\\
& \mathcal{D}_2 (\alpha_p, |\bm{p}_{pT}|^2 \rightarrow - a_T^2)
\; = \; 1 \; + \; \frac{(\alpha_p - 1)^2}{2 (2 - \alpha_p)} + \mathcal{O}\left(\frac{\epsilon_D}{m}\right) .
\label{factor_onshell}
\end{align}
Eq.~(\ref{factor_onshell}) is the deuteron structure factor Eq.~(\ref{D_2}) extrapolated to 
$|\bm{p}_{pT}|^2 = -a_T^2$; the extrapolation is performed using the explicit expressions for 
the spectral functions Eqs.~(\ref{P_unpol})--(\ref{C_T}). A similar expression can be derived for the 
three-state asymmetry. 

In practice, the pole extrapolation can be performed by measuring the
tagged deuteron asymmetry over a range $|\bm{p}_{pT}| \lesssim 100$ MeV and extrapolating the
data using low-order polynomial fits \cite{Sargsian:2005rm}. It is convenient to use the 
two-state rather than the three-state asymmetry, as the former exhibits a much weaker dependence 
on the proton momentum at small momenta (cf.\ Sec.~\ref{sec:deuteron_structure}).
The two-state asymmetry also has the advantage that only preparation
of the $\pm 1$ deuteron spin states is required, reducing the systematic uncertainty.
First simulations of the asymmetry measurements with projected experimental uncertainties at EIC
have been described in Ref.~\cite{Cosyn:2014zfa}.

For a full assessment of the feasibility of the pole extrapolation one should estimate also the effects 
of final-state interactions between the tagged proton and the DIS products. Such interactions can modify 
the tagged DIS cross section at physical proton momenta but are suppressed at the pole; they can thus
affect the practical performance of the pole extrapolation but not its theoretical result.
Detailed studies have shown that in the unpolarized tagged cross section, Eq.~(\ref{sigma_3}),
final-state interaction corrections to the impulse approximation are moderate at momenta 
$|\bm{p}_{pT}| \lesssim 100$ MeV and do not impede the pole extrapolation \cite{Strikman:2017koc}.
The longitudinally polarized cross section, Eq.~(\ref{sigma_polarized}), is similar to the 
unpolarized one in the sense that in both cases the impulse approximation is given by the square 
of the dominant S-wave amplitude. One can therefore expect that the final-state interaction effects 
at momenta $|\bm{p}_{pT}| \lesssim 100$ MeV are not substantially larger in the longitudinally 
polarized cross section than in the unpolarized one. Quantitative estimates could be obtained
by generalizing the methods of Ref.~\cite{Strikman:2017koc}, but require extensive modeling 
of the dynamical input (polarized neutron fragmentation, spin-dependent rescattering).
\section{Conclusions and extensions}
The main conclusions of the present study can be summarized as follows:
(a)~The measured proton momentum in tagged DIS effectively controls the spin structure 
of the $pn$ configuration in deuteron. This feature can be used to select pure S-wave 
configurations and eliminate D-wave depolarization. (b)~The two-state longitudinal spin asymmetry 
has simpler properties than the three-state asymmetry in tagged DIS. 
In the two-state asymmetry the nuclear structure factor is bounded by unity,
and equal to unity at $|\bm{p}_{pT}| = 0$, and the D-wave contributions vanish
rapidly at small proton momenta.
(c)~The free neutron spin asymmetry can be extracted by pole extrapolation of the tagged 
spin asymmetries in $|\bm{p}_{pT}|^2$. 
The method effectively accesses non-interacting large-size $pn$ configurations
through analytic continuation. The extrapolation of the asymmetries is technically simple because 
the pole factors cancel between numerator and denominator.
In sum, DIS on the polarized deuteron with proton tagging permits control of the $pn$ configuration 
during the polarized DIS process (momentum, spin, interactions) and enables new ways of neutron 
spin structure extraction.

The theoretical methods for high-energy scattering on the polarized deuteron with spectator tagging
described here can be applied and extended to several other types of measurements of interest \cite{CW}. 
This includes tagged DIS with transverse deuteron polarization; azimuthal asymmetries
and tensor-polarized tagged structure functions; tagged measurements of hard exclusive processes 
on the neutron such as deeply-virtual Compton scattering; and the use of tagging for studies
of nuclear modifications of partonic structure.

This material is based upon work supported by the U.S.~Department of Energy, 
Office of Science, Office of Nuclear Physics under contract DE-AC05-06OR23177.
It is partly based upon work supported by Jefferson Lab's 2014/2015 Laboratory-Directed 
Research and Development Project ``Physics potential of polarized light ions with EIC@JLab'' 
(see Ref.~\cite{LD1506}). W.C.~acknowledges the hospitality of the Jefferson Lab Theory Center.

\bibliographystyle{elsarticle-num}
\bibliography{polvec.bib}
\end{document}